\documentclass[floatfix,tightenlines,aps,prd,lengthcheck,sort&compress, showpacs,nofootinbib,superscriptaddress]{revtex4-1}

\usepackage{amsmath}
\usepackage{amssymb}
\usepackage{graphicx}
\usepackage{dsfont}
\usepackage{dcolumn}
\usepackage{units}
\usepackage{wasysym}
\usepackage{multirow}
\usepackage{slashed}
\usepackage[usenames]{color}
\usepackage[
	colorlinks=true,
	pdfencoding=auto,
	psdextra,
	linkcolor=blue,
	citecolor=blue,
	urlcolor=blue
	]{hyperref}

\usepackage{slashed}
\usepackage{booktabs}
\usepackage{wasysym}

\definecolor{violet}{RGB}{111,0,255}
\definecolor{dgreen}{rgb}{0.0,0.50,0.0}
\definecolor{webred}{rgb}{0.75,0,0}
\definecolor{fuchsia}{RGB}{255,0,255}

\begin{document}

\title{Electromagnetic decays of the neutral pion}

\author{Esther Weil}
\email[e-mail: ]{Esther.D.Weil@physik.uni-giessen.de}
\affiliation{Institut f\"ur Theoretische Physik, Justus-Liebig Universit\"at Gie{\ss}en, 35392 Gie{\ss}en, Germany}

\author{Gernot Eichmann}
\email[e-mail: ]{Gernot.Eichmann@tecnico.ulisboa.pt}
\affiliation{Institut f\"ur Theoretische Physik, Justus-Liebig Universit\"at Gie{\ss}en, 35392 Gie{\ss}en, Germany}
\affiliation{CFTP, Instituto Superior T\'ecnico, Universidade de Lisboa, 1049-001 Lisboa, Portugal}

\author{Christian S. Fischer}
\email[e-mail: ]{Christian.Fischer@physik.uni-giessen.de}
\affiliation{Institut f\"ur Theoretische Physik, Justus-Liebig Universit\"at Gie{\ss}en, 35392 Gie{\ss}en, Germany}
\affiliation{HIC for FAIR Gie{\ss}en, 35392 Gie{\ss}en, Germany}

\author{Richard Williams}
\email[e-mail: ]{Richard.Williams@physik.uni-giessen.de}
\affiliation{Institut f\"ur Theoretische Physik, Justus-Liebig Universit\"at Gie{\ss}en, 35392 Gie{\ss}en, Germany}

\begin{abstract}
We complement studies of the neutral pion transition form factor  $\pi^0 \rightarrow \gamma^{(*)} \gamma^{(*)}$ with calculations for the electromagnetic decay widths of the processes $\pi^0 \rightarrow e^+ e^-$, $\pi^0 \rightarrow e^+ e^- \gamma$ and $\pi^0 \rightarrow e^+ e^- e^+ e^-$. Their common feature is that the singly- or doubly-virtual transition form factor serves as a vital input that is tested in the non-perturbative low-momentum region of QCD. We determine this form factor from a well-established and symmetry-preserving truncation of the Dyson-Schwinger equations.
Our results for the three- and four-body decays match results of previous theoretical calculations and experimental measurements.
For the rare decay we employ a numerical method to calculate the process directly by deforming integration contours, which in principle can be generalized to arbitrary integrals as long as the analytic structure of the integrands are known. Our result for the rare decay is in agreement with dispersive calculations but still leaves a $2\sigma$ discrepancy between theory and experiment.
\end{abstract}


\maketitle

\section{Introduction}
Over the last years, low-energy electromagnetic processes in the meson sector have seen continuous interest both in the theoretical and experimental physics communities.
Electromagnetic decays of the pion are particularly interesting because they combine the non-perturbative physics
of dynamical mass generation and the associated generation of (pseudo-) Goldstone bosons with the Abelian anomaly and its perturbative elements,
 thus creating an interesting laboratory for theoretical approaches to non-perturbative QCD.
 Moreover, the rare decay $\pi^0 \to e^+ e^-$ poses a puzzle since theoretical estimates show a discrepancy
 with the experimental result from the KTeV E799-II experiment at Fermilab~\cite{Abouzaid:2006kk,Dorokhov:2007bd} of similar magnitude as the muon g-2.

In this work we focus on electromagnetic decays of pseudoscalar mesons such as the rare decay $\pi^0 \to e^+ e^-$,
the Dalitz decay $\pi^0 \to e^+ e^- \gamma$ and double Dalitz decay $\pi^0 \to e^+ e^- e^+ e^-$
using a well explored combination of Dyson-Schwinger and Bethe-Salpeter equations.
All calculations involve the $\pi^0\to\gamma^{(\ast)}\gamma^{(\ast)}$ transition form factor (TFF) with one or two off-shell photons, thus testing it
in the region of (very) low momenta. The present work complements a recent evaluation of the TFF
at large space-like momenta, see Ref~\cite{Eichmann:2017wil}.

The paper is organised as follows. In the next section we give a short introduction to the details of our calculations
and discuss features of the resulting TFF. In section~\ref{sec3} we then give results for the leptonic
three- and four-body decays of the neutral pion and compare with the experimental values. In section~\ref{sec4} we discuss
corresponding results for the rare decay of the pion into an electron-positron pair. We conclude in section~\ref{sec5}.

\section{Transition form factor \texorpdfstring{$\pi^0 \to \gamma^\ast \gamma^\ast$}{\pi0 \to \gamma\star \gamma\star}}\label{sec2}
\subsection{Kinematics and definitions}\label{sec:kinematics}

We begin by defining the transition current and the kinematic regions of interest. The $\pi^0\to\gamma\gamma$ transition matrix element is given by
\begin{align}\label{pigg-current}
	\Lambda^{\mu\nu}(Q,Q') = e^2\,\frac{F(Q^2,{Q'}^2)}{4\pi^2 f_\pi}\,\varepsilon^{\mu\nu\alpha\beta}  {Q'}^\alpha Q^\beta \,,
\end{align}
where $Q$ and $Q'$ are the photon momenta, $f_\pi\approx 92$~MeV is the pion's electroweak decay constant,  and $e^2=4\pi \alpha_\text{em}$ the squared electromagnetic charge.
The pseudoscalar transition is described by a single scalar function, the transition form factor $F(Q^2,{Q'}^2)$,
and the convention of prefactors is such that $F(0,0)=1$ in the chiral limit due to the Abelian anomaly.

In the following it is useful to work with the average photon momentum $\Sigma$ and the pion momentum $\Delta$,
\begin{align}\label{kinematics-0}
\Sigma = \frac{Q+Q'}{2}\,, \qquad
\Delta = Q-Q'\,,
\end{align}
with $\Delta^2  = -m_\pi^2$ for an on-shell pion.
The process then depends on two Lorentz invariants,
\begin{align}\label{li-2}   \renewcommand{\arraystretch}{1.2}
\eta_+ &=  \displaystyle \frac{Q^2+{Q'}^2}{2}  = \Sigma^2+\frac{\Delta^2}{4}\,, \\
\omega &= \displaystyle \frac{Q^2-{Q'}^2}{2}   = \Sigma\cdot \Delta\,,
\end{align}
or vice versa: $\{ Q^2, \, {Q'}^2 \} = \eta_+ \pm \omega$, with the third invariant fixed when the pion is on-shell:
\begin{align}
  \eta_- =  Q\cdot Q'  = \Sigma^2-\frac{\Delta^2}{4} = \eta_+ + \frac{m_\pi^2}{2} \,.
\end{align}
Note that the TFF is symmetric in $Q^2$ and ${Q'}^2$ so it can only depend on $\omega$ quadratically.

For practical calculations it is convenient to introduce the alternative variables
\begin{align}\label{alt-kinematics}
  \sigma = \Sigma^2\,, \quad t = \frac{\Delta^2}{4}\,, \quad Z = \hat{\Sigma}\cdot\hat{\Delta} = \frac{\Sigma\cdot\Delta}{2\sqrt{\sigma t}}
\end{align}
where $t=-m_\pi^2/4$ when the pion is on-shell and a hat denotes a normalized four-vector.
We will refer to those in section~\ref{sec:rare-decay} and Appendix~\ref{app:sing}.

\begin{figure}[!t]
\begin{center}
\includegraphics[width=0.87\columnwidth]{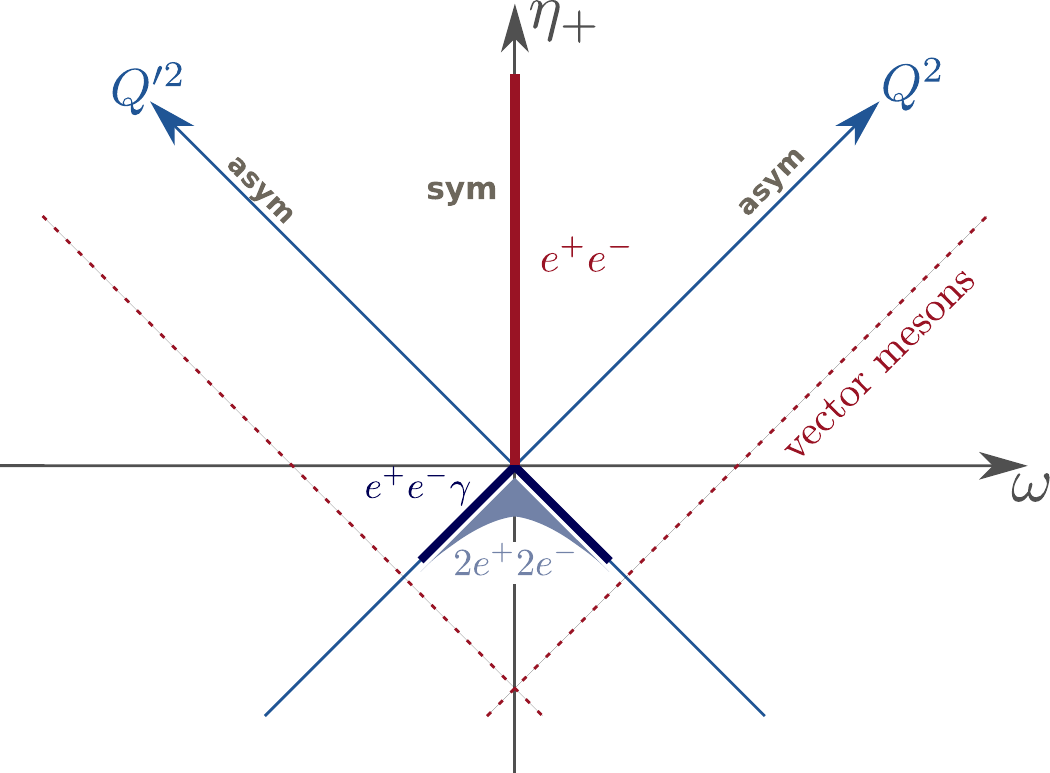}
\caption{Kinematic domains in $Q^2$ and $Q'^2$ including the symmetric (red line) and asymmetric limits ($Q^2$, $Q'^2$ axes).
The area with $|\omega|< \eta_+$ corresponds to space-like momentum transfer. In the timelike region we show the relevant domains
for the Dalitz and double Dalitz decays and the dotted lines indicate the vector-meson pole locations  (not to scale).\label{fig:phasespace-1}}
\end{center}
\end{figure}

In the physical processes we study in this work the TFF is tested in both space-like and time-like regions as shown in Fig.~\ref{fig:phasespace-1}.
The time-like region, where either $Q^2<0$ or ${Q'}^2<0$, contains the physical singularities such as the vector-meson poles in the complex plane of $Q^2$ and ${Q'}^2$. For dilepton decays this region is kinematically restricted such that $Q^2,Q'^2\ge -m_\pi^2$. The double Dalitz decay $\pi\to 2 e^+ 2e ^-$ is constrained to the light blue shaded area below $\eta_+<0$, whilst the Dalitz decay $\pi\to  e^+ e^- \gamma$ probes the asymmetric time-like form factor, given by the dark blue lines along the $Q^2,Q'^2$ axes. These decays are discussed in section~\ref{sec3}.

 The space-like region with both photon virtualities positive, $Q^2>0$ and $Q'^2>0$, is free of any physical singularities.
 The region that is relevant for the rare decay $\pi^0 \to e^+ e^-$, discussed in section~\ref{sec4}, is the
 doubly-virtual or symmetric limit (red line in  Fig.~\ref{fig:phasespace-1}) when $Q^2=Q'^2$ viz. $\omega=0$.
Direct experimental measurements of the spacelike TFF are available in the
singly-virtual or asymmetric limit with one of $\left\{Q^2,Q'^2\right\}$ vanishing~\cite{Behrend:1990sr,Gronberg:1997fj,Aubert:2009mc,Uehara:2012ag}.

In addition, very different kinematic regions to these can be tested where the pion is `off-shell' corresponding to space-like momentum transfer $\Delta^2>0$. This and various applications are discussed in~\cite{Eichmann:2017wil}.

\subsection{Triangle diagram}

\begin{figure}[!b]
\begin{center}
\includegraphics[width=0.7\columnwidth]{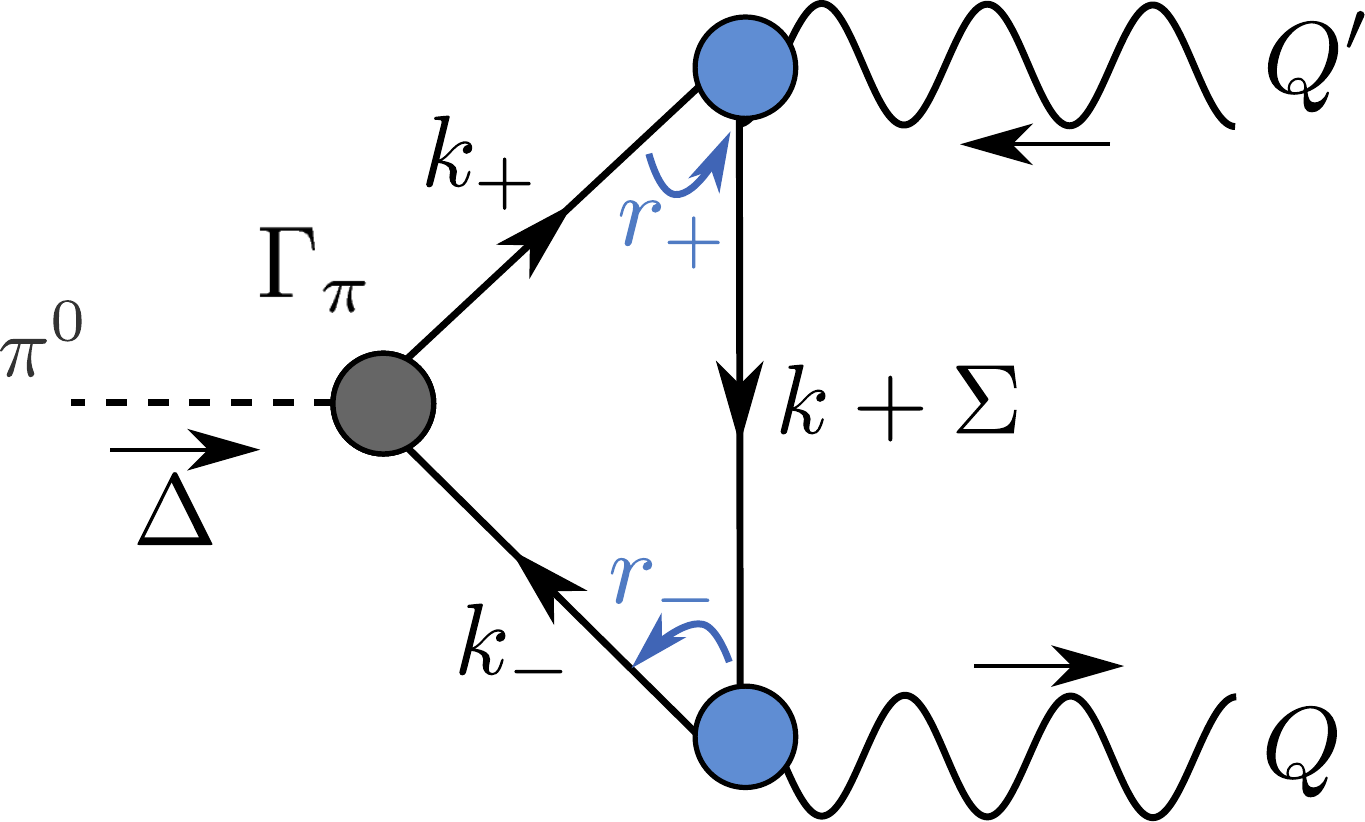}
\caption{The transition form factor given by Eq.~\eqref{eqn:PseudoScalarFormFactor}. The non-perturbative input is the Bethe-Salpeter amplitude $\Gamma_\pi$ of the pion (grey circle), the dressed quark propagators (straight lines) and the dressed quark-photon vertices $\Gamma_\nu$ (blue circles). \label{fig.pigg:diagram} }
\end{center}
\end{figure}
In the impulse approximation, the transition form factor $\pi^0 \rightarrow \gamma^{(*)} \gamma^{(*)}$ is displayed in Fig.~\ref{fig.pigg:diagram} and given by
\begin{align} \label{eqn:PseudoScalarFormFactor}
\Lambda^{\mu\nu} &= 2e^2\, \text{Tr} \int \!\! \frac{d^4k}{(2\pi)^4} \,  S(k_+)\,\Gamma_\pi(k,\Delta)\,S(k_-) \nonumber  \\
       &\times \Gamma^\mu(r_-,-Q)\,S(k+\Sigma)\,\Gamma^\nu(r_+,Q')\,.
\end{align}
The photon momenta were defined in the previous subsection. In addition, $k$ is the loop momentum and
\begin{align}\label{kin-1}
k_\pm = k \pm \frac{\Delta}{2}\,, \qquad
r_\pm = k + \frac{\Sigma}{2} \pm \frac{\Delta}{4}\;,
\end{align}
are the internal quark momenta and the relative momenta appearing in the quark-photon vertices, respectively.
The trace in Eq.~\eqref{eqn:PseudoScalarFormFactor} is over Dirac indices only\footnote{All quantities are color-diagonal and thus the color trace is 3. The flavor matrix of the $\pi^0$ amplitude is $\mathrm{diag}\,(1,-1)$ and that of the quark-photon vertex is $\mathrm{diag}\,(q_u,q_d)$, so the flavor trace is $q_u^2-q_d^2 = 1/3$. Since the overall normalization of the pion amplitude is determined by the canonical Bethe-Salpeter norm, which follows from demanding unit residue at the pion pole in the $q\bar{q}$ scattering matrix, a different color-flavor normalization can always be absorbed by the dressing functions in Eq.~\eqref{pion-amplitude}. Our choice above is such that  $f_1(k^2) = B(k^2)/f_\pi$ in the chiral limit, with $B(k^2) = M(k^2)/Z_f(k^2)$.}
and the factor 2 in front of the integral stems from the exchange of the photons.

All ingredients in Eq.~\eqref{eqn:PseudoScalarFormFactor} are determined from numerical solutions of their Dyson-Schwinger and Bethe-Salpeter equations. The renormalized quark propagator is given by
\begin{align}\label{quarkprop}
S(p) = Z_f(p^2)\,\frac{-i \slashed{p}  + M(p^2)}{p^2  + M^2(p^2)}
\end{align}
with non-perturbative dressing functions $Z_f(p^2)$ and $M(p^2)$. The renormalization-group invariant quark mass function $M(p^2)$ encodes effects of dynamical mass generation due to the dynamical breaking of chiral symmetry. The Dirac structure of the pseudoscalar Bethe-Salpeter amplitude $\Gamma_\pi$ is given by
\begin{align}\label{pion-amplitude}
\Gamma_\pi(k,\Delta) =  \left(f_1 + f_2\,i\slashed{\Delta}  + f_3\, k\cdot \Delta \,i\slashed{k} +f_4 \left[\slashed{k},\slashed{\Delta}\right]\right) \gamma_5,
\end{align}
where the $f_i$ are functions of $k^2$ and $k\cdot \Delta$ with $\Delta^2=-m_\pi^2$ fixed.

The non-perturbative quark-photon vertex $\Gamma^\mu$ describes the coupling of a dressed quark to a photon and is dominated by QCD corrections. It can be decomposed into twelve tensor structures,
\begin{align}
\Gamma^\mu(p,Q) = \sum_{i=1}^{12} \lambda_i(p^2,p\cdot Q, Q^2) \,\tau_i^\mu(p,Q) \label{vertex}
\end{align}
with basis components $\tau_i^\mu(p,Q)$ and Lorentz-invariant dressing functions $\lambda_i$; see App.~B of Ref.~\cite{Eichmann:2016yit} for details. The argument $p$ denotes the average momentum of the two quark legs and $Q$ is the incoming photon momentum. Due to electromagnetic gauge invariance the vertex can be split into a transverse and non-transverse part, where the latter is the Ball-Chiu vertex~\cite{Ball:1980ay} and determined by the vector Ward-Takahashi identity. We obtain a numerical solution of the quark-photon vertex from its inhomogeneous Bethe-Salpeter equation; see~\cite{Maris:1999bh,Maris:1999ta,Maris:2002mz,Bhagwat:2006pu,Goecke:2010if,Eichmann:2011vu}. As discussed in detail in~\cite{Maris:1999bh,Goecke:2012qm,Eichmann:2016yit}, the transverse part of the vertex contains poles in the time-like momentum region corresponding to vector-meson states. Thus, the underlying physics of vector-meson dominance is automatically contained in the numerical representation of the vertex without the need for further adjustments.

In this work we restrict ourselves to the rainbow-ladder approach as reviewed in~\cite{Eichmann:2016yit}.
We use the Maris-Tandy model, Eq.~(10) of Ref.~\cite{Maris:1999nt} with parameters $\Lambda=0.74$ GeV and $\eta = 1.85 \pm 0.2$
(the parameters $\omega$ and $D$ therein are related to the above via $\omega D = \Lambda^3$ and $\omega=\Lambda/\eta$).
The variation of $\eta$ changes the shape of the quark-gluon interaction in the infrared, cf.~Fig. 3.13 in Ref.~\cite{Eichmann:2016yit},
and we use it in the following to estimate our theoretical error.
This construction, with the same respective kernel in the Bethe-Salpeter equation for the pseudoscalar mesons and the one for the quark-photon vertex,
preserves chiral symmetry and has the merit of producing reliable results in the pseudoscalar and vector-meson sector as well as for nucleon and $\Delta$ baryons.
Our input current-quark mass is $m_q=3.57$ MeV at a renormalization point $\mu=19$~GeV; the resulting pion mass and pion decay constant are $m_{\pi^0} = 135.0(2)$ MeV and $f_{\pi^0} = 92.4(2)$ MeV.

\begin{figure}[!t]
\begin{center}
\includegraphics[width=0.7\columnwidth]{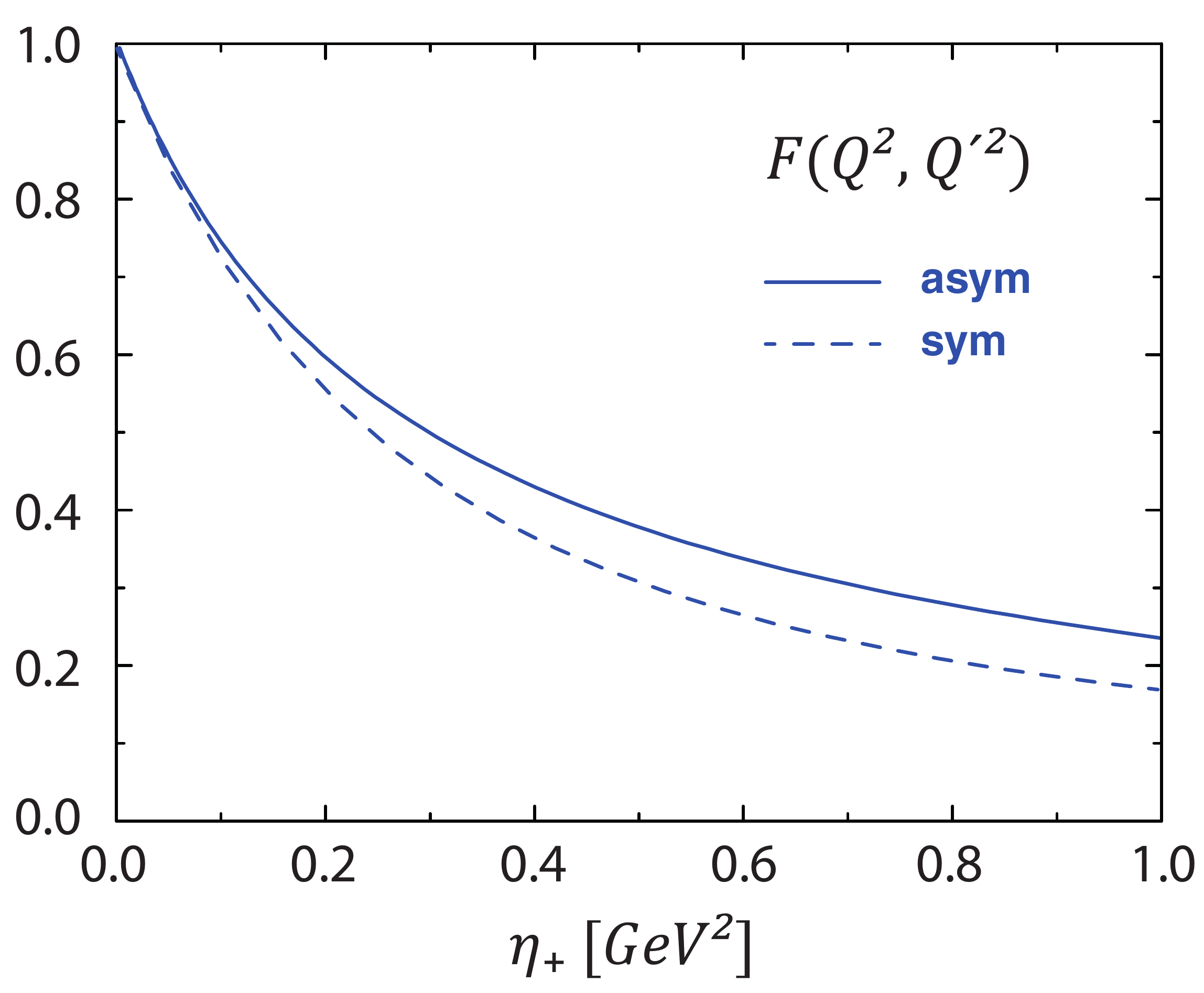}
\caption{On-shell transition form factor in the symmetric limit ($\eta_+=Q^2=Q'^2$) and asymmetric limit ($\eta_+=Q^2/2$, ${Q'}^2=0$)
         for moderate space-like values $\eta_+>0$.\label{fig:ff-lowQ2}}
\end{center}
\end{figure}

\subsection{Result for the transition form factor}\label{sec:tff-result}
With the ingredients described above we are able to determine the $\pi^0\to\gamma\gamma$ transition form factor in the spacelike domain $Q^2>0$ and ${Q'}^2>0$
as well as for small timelike momenta.
In practice it turns out that a straightforward calculation is only possible in restricted kinematic regions. As explained in Appendix~\ref{app:sing},
this is due to the singularities of the quark propagator in the integrand, whose nearest singularities correspond to a scale $m_p \sim 0.5$ GeV.
The symmetric limit is accessible for all $\eta_+>0$, whereas in the asymmetric limit one is limited to $Q^2_\text{max} \approx 4$ GeV$^2$, which is also the domain
covered in the calculation of Ref.~\cite{Maris:2002mz}. In addition, also small timelike momenta of the order $Q^2$, ${Q'}^2 \gtrsim -m_p^2$ are accessible directly;
cf.~Fig.~\ref{fig:sing} in the appendix.

To determine the TFF in the full spacelike domain,
we employ the strategy introduced in Ref.~\cite{Eichmann:2017wil}: we calculate the form factor for an off-shell pion with $\Delta^2>0$ and extrapolate to the on-shell point
using the lowest-lying vector-meson pole as a constraint. This allows us to determine the TFF for all space-like momenta, and
in the regions that are accessible by direct calculation and extrapolation we have checked that the results of both methods are in perfect agreement.

\begin{figure*}[!t]
\begin{center}
\includegraphics[width=0.95\textwidth]{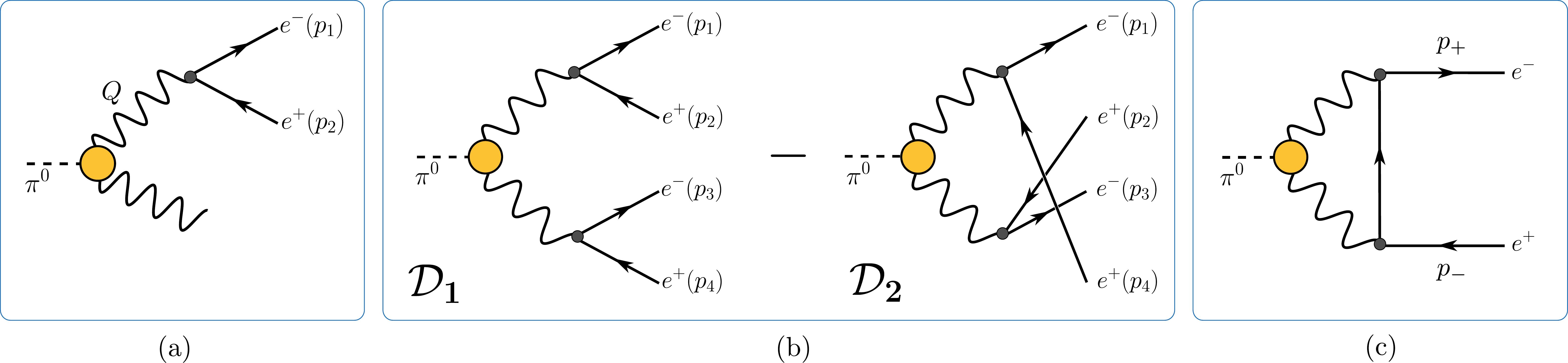}
\caption{The tree-level contributions to the (a) single Dalitz decay,
         (b) double Dalitz decay and (c) rare decay of the neutral pion. The yellow circle denotes the
         pion transition form factor $F(Q^2,Q'^2)$, see Eq.~\eqref{pigg-current}.\label{fig:alldecays}}
\end{center}
\end{figure*}

In Fig.~\ref{fig:ff-lowQ2} we show the on-shell transition form factor $F(Q^2,{Q'}^2)$
as a function of the variable $\eta_+$, Eq.~\eqref{li-2}. The plot reveals that the TFF is essentially a function of $\eta_+$, which is larger in the
asymmetric limit and smaller in the symmetric limit, and this behavior persists up to asymptotically large $\eta_+$~\cite{Eichmann:2017wil}.
Moreover, in the chiral limit the Abelian anomaly entails $F(0,0)=1$ and
our numerical result at the physical pion mass is $F(0,0) = 0.996$, which
provides an important consistency check:
replacing both dressed quark-photon vertices by bare ones would only give $F(0,0) \approx 0.29$;
and even a Ball-Chiu vertex, which guarantees charge conservation in the pion's electromagnetic form factor,
produces $F(0,0) \approx 0.86$ only. The transverse structure of the vertex is therefore crucial for a quantitative description of the $\pi^0\to\gamma\gamma$ transition.

We finally provide a fit function that accurately represents our results in the spacelike region.
Abbreviating $w=\eta_+/m_v^2$ and $z=\omega/\eta_+$, the TFF is described by
\begin{equation}
 F(Q^2,{Q'}^2) = \frac{\mathcal{A}(w) + w(1-z^2)\,\mathcal{B}_1(w)\,(1+\mathcal{B}_2(w) z^2)} {(1+w)^2-w^2 z^2}\,.
\end{equation}
The denominator implements the lowest-lying vector-meson pole at $wz = \pm(1+w)$, which corresponds to $Q^2=-m_v^2$ and ${Q'}^2 = -m_v^2$ with $m_v=0.77$ GeV.
The functions in the numerator ensure that the TFF asymptotically approaches a monopole behavior both in the symmetric ($z=0$) and asymmetric limit ($z=\pm 1$);
they are given by
\begin{equation}\label{fit}
\begin{split}
  \mathcal{A}(w) &= \frac{a_0+\xi\,(a_1\, b_1\, w+a_2 \,b_2 \,w^2+a_3 \,b_3\,w^3)}{1+b_1 \,w+b_2 \,w^2+b_3 \,w^3}\,, \\
  \mathcal{B}_i(w) &= \frac{c_i \,e_i \,w^2}{1+d_i \,w + e_i \,w^2}
\end{split}
\end{equation}
with fit parameters $a_0=0.996$ and
\begin{equation}\label{fit-parameters}  \renewcommand{\arraystretch}{1.1}
\begin{array}{rl}
  a_1 &= 0.735\,,\\
  a_2 &= 1.214\,,\\
  a_3 &= 1.547\,, \\[1mm]
  c_1 &= 0.384\,,\\
  d_1 &= 2.010\,,\\
  e_1 &= 1.540\,,
\end{array}\qquad
\begin{array}{rl}
  b_1 &= 0.089\,,\\
  b_2 &= 0.133\,,\\
  b_3 &= 0.0002\,,  \\[1mm]
  c_2 &= 0.430\,,\\
  d_2 &= 0.024\,,\\
  e_2 &= 0.00005\,.
\end{array}
\end{equation}
This fit provides the input for our calculations of the various $\pi^0$ decays.
The value $\xi = 1.0 \pm 0.1$ reflects our combined theoretical uncertainty from varying the parameter $\eta = 1.85 \pm 0.2$ in the effective interaction
as well as the uncertainty in the determination of the TFF away from the symmetric limit. These are also the error estimates that we quote in the following results.

Let us  finally briefly comment on alternative fit functions available in the literature (see, e.g., Appendix~B of Ref.~\cite{Nyffeler:2016gnb} for a discussion).
The simplest vector-meson dominance (VMD) parametrization $F_\text{VMD}(Q^2,{Q'}^2) = 1/[(1+w)^2-w^2 z^2]$ does not reproduce the monopole behavior in the symmetric limit $z=0$
but instead approaches a dipole at large $Q^2$. Its refined version based on lowest-meson dominance, the LMD+V model~\cite{Knecht:2001xc},
reproduces both the symmetric and the asymmetric limits and implements two vector-meson poles; an analogous form was recently employed to fit lattice results for the TFF~\cite{Gerardin:2016cqj}.
Our fit is practically indistinguishable from the LMD+V parametrization at low $Q^2$,
i.e., in the momentum range shown in Fig.~\ref{fig:ff-lowQ2}. Also at large $Q^2$ the fits in the symmetric limit are almost identical  and in the asymmetric limit they are at least qualitatively similar.
However, the behavior in between ($0<|z|<1$) differs
substantially: at large~$\eta_+$ the LMD+V form factor develops a sharp peak very close to $|z|=1$, with a turnover to reach the asymmetric point $z=1$ followed by the vector-meson poles at $|z|>1$.
By comparison, our fit varies monotonously from $z=0$ to $z=1$ and is therefore better suited for applications where the TFF is tested in the whole spacelike domain.

\section{Three- and four-body decays}\label{sec:decays}\label{sec3}
In this subsection we discuss our results for the three- and four-body decays of pseudoscalar mesons
shown in Fig.~\ref{fig:alldecays}(a-b).
The Dalitz decay of the neutral pion into a photon and an electron-positron pair
has the largest branching ratio $B(\pi^0\to e^+ e^-\gamma)= (1.174 \pm 0.035) \%$ \cite{Patrignani:2016} after that of the two-photon decay.
The neutral pion also decays into two dilepton pairs with a branching ratio of
$B(\pi^0\to e^+ e^-e^+e^-)= (3.34 \pm 0.16)\times 10^{-5}$ \cite{Patrignani:2016}.
Both decays depend on the transition form factor discussed above as the only non-trivial input.

\subsection{Dalitz Decay: \texorpdfstring{$\pi^0\to e^+ e^- \gamma$}{\pi0 \rightarrow e+ e- \gamma}}

The leading-order Feynman diagram for the three-body decay of the neutral pion is shown in Fig.~\ref{fig:alldecays}(a).
The decay rate is easily calculated and given by
\begin{align}\label{dalitz-eq}
& \Gamma_{\pi^0\rightarrow e^+e^- \gamma} =  \frac{e^6 m_\pi^3}{6(4 \pi)^3}\int_{4 m^2}^{m_\pi^2}    \frac{dx}{x} \left|  \frac{F(Q^2,0)}{4\pi^2f_\pi} \right|^2  \nonumber \\
& \qquad \times \sqrt{1-\frac{4 m^2}{x}} \left(1 + \frac{2 m^2}{x}\right) \left(1 - \frac{x}{m_\pi^2}\right)^3,
\end{align}
where $m$ is the electron mass, $m_\pi$ the pion mass, and
$x=-Q^2$ is the momentum squared of the virtual photon which evaluates the form factor in the timelike region.

%
\begin{figure}[!b]
\begin{center}
\includegraphics[width=0.95\linewidth]{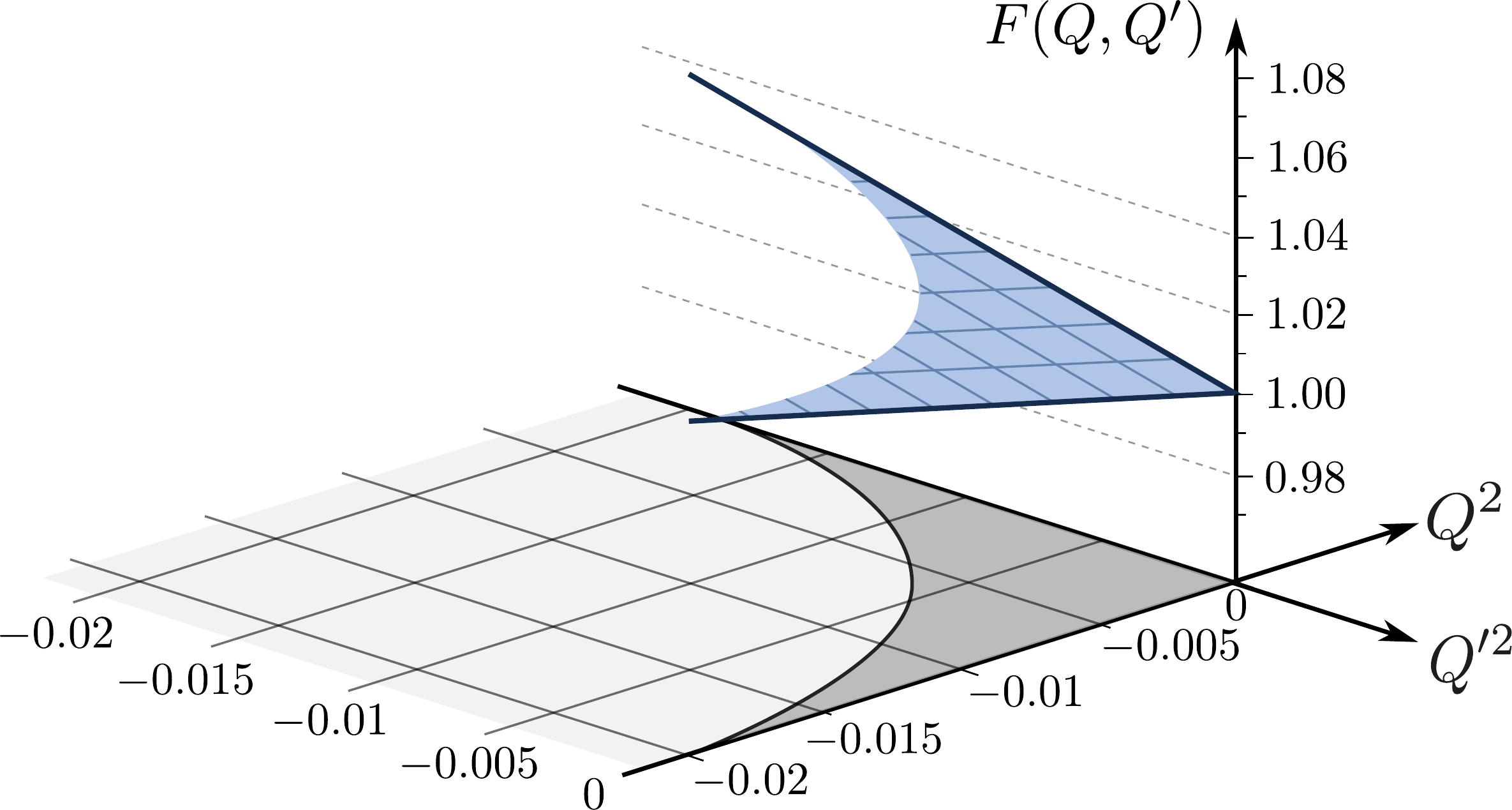}
\caption{Transition form factor shown in the region relevant for the three and four-body decays. The singly-virtual form factor
         for the three-body decay, for which either $Q^2=0$ or $Q'^2=0$, is indicated by the heavy blue lines.\label{fig.gepem:phasespace}}
\end{center}
\end{figure}

Due to the kinematics of the three-body decay the TFF is probed in the asymmetric region of one photon on-shell and one off-shell and time-like.
The leading contribution to the integral in Eq.~\eqref{dalitz-eq} comes from the lower end of the integral.
In this region the calculated TFF can in general be described by a simple linear fit with respect to $\eta_+= (Q^2+Q'^2)/2$,
\begin{align}\label{linear-fit}
F(Q^2, Q'^2) = 0.996 -3.55(10)\,\eta_+  \;,
\end{align}
which is also used in the four-body decay below. The TFF is shown in Fig.~\ref{fig.gepem:phasespace},
where the asymmetric limit required for the three-body decay, $F(Q^2,0)$, is represented by either one of the dark blue lines.

Employing the PDG values for $m$ and $\alpha_\text{em}$ in Eq.~\eqref{dalitz-eq},
our calculated decay width is $\Gamma_{\pi\to   e^+ e^-\gamma}=9.11(4)\times10^{-11}$~GeV.
Since the theoretical uncertainty in Eq.~\eqref{linear-fit}  affects this number only at the sub-per-mille level,
the error bar in the decay rate comes from the model dependence of $m_\pi$ and $f_\pi$ which enter in Eq.~\eqref{dalitz-eq}.
In Table~\ref{tab:epemgs} we compare our result with the PDG
and the result of a calculation using an effective theory \cite{Terschlusen:2013iqa}.
Within the quoted errors all results are in good agreement. This is to be expected because the TFF -- as the only non-perturbative input --
is probed in the kinematic region governed by the anomaly. Since any reasonable construction obeys this constraint,
the discriminative potential of the Dalitz decay with regard to different non-perturbative input is very limited.

\begin{table}[!h]
\begin{center}
\begin{tabular}{cl}
\toprule
Collaboration 								& $\Gamma_{\pi^0\to   e^+ e^-\gamma}\, [10^{-11}$ GeV] \\
\midrule \hline
PDG~\cite{Patrignani:2016} 				   	    &\hspace{1cm} $9.06 (18)$\\
Terschl\"usen et al.~\cite{Terschlusen:2013iqa} &\hspace{1cm} $9.26$\\
Hoferichter et al.~\cite{Hoferichter:2014vra}   &\hspace{1cm}  $9.065$\\
Our result  								    &\hspace{1cm} $9.11(4)$\\
\bottomrule
\end{tabular}
\caption{Result for the Dalitz decay.\label{tab:epemgs} }
\end{center}
\end{table}

\subsection{Four-body decay: \texorpdfstring{$\pi^0 \rightarrow e^+e^-e^+e^- $}{\pi0 \rightarrow e+ e- e+ e-}}

We now proceed with the four-body decay of the neutral pion into two electron-positron pairs. The decay rate is given by
\begin{align}\label{four-body-decay-rate}
\Gamma_{\pi^0 \rightarrow 2e^+2e^-} &= \frac{1}{(2 !)^2 } \,\frac{1}{2m_\pi} \int d \Phi_4\,  |\mathcal{M}|^2 \, ,
\end{align}
where $\left|\mathcal{M}\right|^2$ is the squared and spin-summed matrix element
and the symmetry factor in front accounts for the two pairs of identical final-state particles.
$d\Phi_4$ is the four-dimensional phase space measure whose detailed derivation is given in Appendix~\ref{app:4bphase}.
Because the amplitude $\mathcal{M}$ with all initial and final particles on-shell depends on five independent variables,
$d\Phi_4$ involves five non-trivial integrations.

Fig.~\ref{fig:alldecays}(b) shows the possible diagrams, $\mathcal{D}_1$ and $\mathcal{D}_2$, where exchange of two leptons (anti-leptons) introduces a relative minus sign between the two contributions. Squaring these we obtain
\begin{align}\label{eq.2ep2em:squarM}
 \left|\mathcal{M}\right|^2 = \left|\mathcal{D}_{1}\right|^2+ \left|\mathcal{D}_{2}\right|^2 +2 \mathrm{Re}[\mathcal{D}_{1} \mathcal{D}_{2}^*]  \,.
\end{align}
As discussed in Refs.~\cite{Terschlusen:2013iqa,Escribano:2015vjz}, the first two terms are equal and can thus be combined. It follows that the decay rate can be decomposed into
\begin{align}
\Gamma_{\pi^0 \to 2e^+ 2e^-} = \Gamma_{\pi^0 \to 2e^+ 2e^-}^{(\text{direct})}  + \Gamma_{\pi^0 \to 2e^+ 2e^-}^{(\text{indirect})} \,,
\end{align}
where the first (direct) contribution comes from the two squared magnitudes and the second (indirect or interference) term comes from the cross-terms.

\begin{table}[b]
\begin{center}
\begin{tabular}{cl}
\toprule
Collaboration 									& $\Gamma_{\pi^0\rightarrow  2 e^+ 2 e^-}\, [10^{-13}$GeV] \\\midrule \hline
PDG~\cite{Patrignani:2016} 						&\hspace{1cm} $2.58(12)$\\
Terschl\"usen et al.~\cite{Terschlusen:2013iqa}  	& \hspace{1cm}  $2.68$\\
Escribano et al.~\cite{Escribano:2015vjz} 		& \hspace{1cm}  $2.62$ \\
Our result  									&\hspace{1cm} $2.63(1)$\\
\bottomrule
\end{tabular}
\caption{Result for the double Dalitz decay.\label{tab:2ep2em} }
\end{center}
\end{table}

Abbreviating $G^{\mu\nu}_{ij} = (i\slashed{p}_i +m) \,\gamma^\mu   \,(i\slashed{p}_j -m) \,\gamma^{\nu}$,
the integrand for the direct contribution is given by
\begin{equation*}
      |\mathcal{D}_{1}|^2+ |\mathcal{D}_{2}|^2= 2 e^4\frac{\Lambda^{\mu\nu}(Q,Q')\,\Lambda^{\alpha\beta}(Q,Q') }{Q^4\,{Q'}^4}\,\text{Tr}\,G_{34}^{\mu\alpha}\,\text{Tr}\,G_{12}^{\nu\beta}\,,
\end{equation*}
where $\Lambda^{\mu\nu}(Q,Q')$ is the $\pi^0\to\gamma\gamma$ transition current defined in Eq.~\eqref{pigg-current}.
Because in this case the integrand only depends on the pairwise sums $Q'=-(p_1+p_2)$ and $Q=p_3+p_4$, the five-dimensional phase-space integral $d\Phi_4$ can be reduced to just two integrations, namely
\begin{align}\label{eq.pi2l:direct}
&\Gamma_{\pi^0 \to 2e^+ 2e^-}^{(\text{direct})} =
\frac{ e^8}{36\,(2\pi)^5 \,m_\pi^3}
\int\limits_{4 m^2}^{(m_\pi-2m)^2}      \hspace{-7pt} dx
\int\limits_{4m^2}^{(m_\pi-\sqrt{x})^2} \hspace{-7pt} dy
\nonumber \\
& \times \sqrt{x-4m^2} \sqrt{y-4m^2} \left[\frac{(x+y-m_\pi^2)^2}{4xy} - 1\right]^{3/2} \nonumber  \\
& \times \left(1+\frac{2m^2}{x}\right) \left(1+\frac{2m^2}{y}\right) \left|\frac{F(Q^2,{Q'}^2)}{4\pi^2f_\pi}\right|^2 .
\end{align}
Here we used the shorthands $x=-Q^2$ and $y=-{Q'}^2$ for the two photon virtualities, where $x,y>0$ are timelike
and restricted by the thresholds for the two two-body decays. The direct contribution constitutes the largest fraction of the decay rate.

In contrast, the interference term depends on all possible four-vector combinations of electron and positron pairs.
Consequently, the phase-space integral cannot be further reduced and its integrand is given by
\begin{align*}
  \mathrm{Re}[\mathcal{D}_{1} \mathcal{D}_{2}^*]   = e^4\,
\frac{\Lambda^{\mu\nu}(Q,Q')\,\Lambda^{\alpha\beta}(K,K') }{Q^2\,{Q'}^2\,K^2\,{K'}^2} \,  \text{Tr} \, G_{32}^{\mu\beta}\,G_{14}^{\nu\alpha}\,,
\end{align*}
where $Q'=-(p_1+p_2)$, $Q=p_3+p_4$ and $K'=-(p_2+p_3)$, $K=p_1+p_4$ are the possible momenta of the virtual photons.
To perform the five-dimensional integral we used various methods, ranging from tensor-product quadrature with a combination of Gauss-Legendre and double exponential rules, to 5-dimensional adaptive cubature as well as standard Monte-Carlo methods~\cite{Hahn:2004fe}; all agreed perfectly.

For the direct and indirect contributions to the decay rate
we obtain
\begin{align}
 \Gamma_{\pi^0 \rightarrow 2e^+ 2e^-}^{(\text{direct})} &=  \phantom{-}2.66(1) \times 10^{-13} \,\text{GeV},\\
 \Gamma_{\pi^0 \rightarrow 2e^+ 2e^-}^{(\text{indirect})} &= -0.03  \times 10^{-13} \,\text{GeV}.
\end{align}
The sum of these values gives our final result, shown in Table~\ref{tab:2ep2em} together with the value from experiment and other theoretical calculations.
In Ref.~\cite{Terschlusen:2013iqa} the authors use an extension of chiral perturbation theory to calculate the form factor, whereas in Ref.~\cite{Escribano:2015vjz} a data driven approach is employed, which is based on the use of rational approximants applied to the experimental data of the $\pi^0, \eta$  and $\eta^\prime$ transition form factors in the space-like region. All results are compatible with experiment within the quoted error.
As with the Dalitz decay, the phase-space restriction of the virtual photons to  time-like momenta between $4m^2$ and $m_\pi^2$ entails that
the sensitivity to the details of the TFF beyond that dictated by the anomaly is rather small and deviates from its nominal value of $1$ by no more than $3$\%.

\section{Rare decay: \texorpdfstring{$\pi^0 \rightarrow e^+ e^-$}{\pi0 \rightarrow e+ e-}}\label{sec:rare-decay}\label{sec4}

Finally we consider the two-body decay of the neutral pion into one electron-positron pair. For the $\pi^0$ this is certainly the most interesting decay due to a discrepancy between the KTeV experimental result and theoretical calculations~\cite{Abouzaid:2006kk,Dorokhov:2007bd,Dorokhov:2009jd,Vasko:2011pi,Husek:2014tna,Masjuan:2015lca} of the order of $2\sigma$. Using the elaborate reanalysis of radiative corrections \cite{Vasko:2011pi,Husek:2014tna} to the experimental result of the KTeV collaboration \cite{Abouzaid:2006kk} (close to the value given in PDG \cite{Patrignani:2016}) one arrives at an extracted experimental value for the branching ratio of $B(\pi^0 \rightarrow e^+ e^-) = (6.87 \pm 0.36) \times 10^{-8}$, which is considerably smaller than the decays considered above.

To lowest order in QED the process is described by the one-loop graph in Fig.~\ref{fig:alldecays}(c), which again includes the transition form factor $F(Q^2,{Q'}^2)$ as the only non-perturbative input.
Defining $t = \Delta^2/4$ as in Eq.~\eqref{alt-kinematics}, the corresponding normalized branching ratio is given by
\begin{align}
R = \frac{B(\pi^0\rightarrow e^+ e^- )}{B(\pi^0\rightarrow \gamma\gamma )}  = 2 \left(\frac{m\,\alpha_\text{em}}{\pi m_\pi}\right)^2 \beta(t_0) \  \vert \mathcal{A}(t_0)\vert^2\,,
\end{align}
where $\beta(t)= \sqrt{1+ m^2 /t}$ stems from the two-body phase-space integration and $B(\pi^0\to \gamma\gamma )=0.988$. The scalar amplitude $\mathcal{A}(t)$ can be viewed as the pseudoscalar form factor of the electron due to the two-photon coupling, which must be evaluated at the on-shell pion point $t_0 = -m_\pi^2/4$.

\subsection{\texorpdfstring{$\mathcal{A}(t)$}{A(t)} with dispersive input}

\begin{figure}[!t]
\begin{center}
\includegraphics[width=0.65\columnwidth]{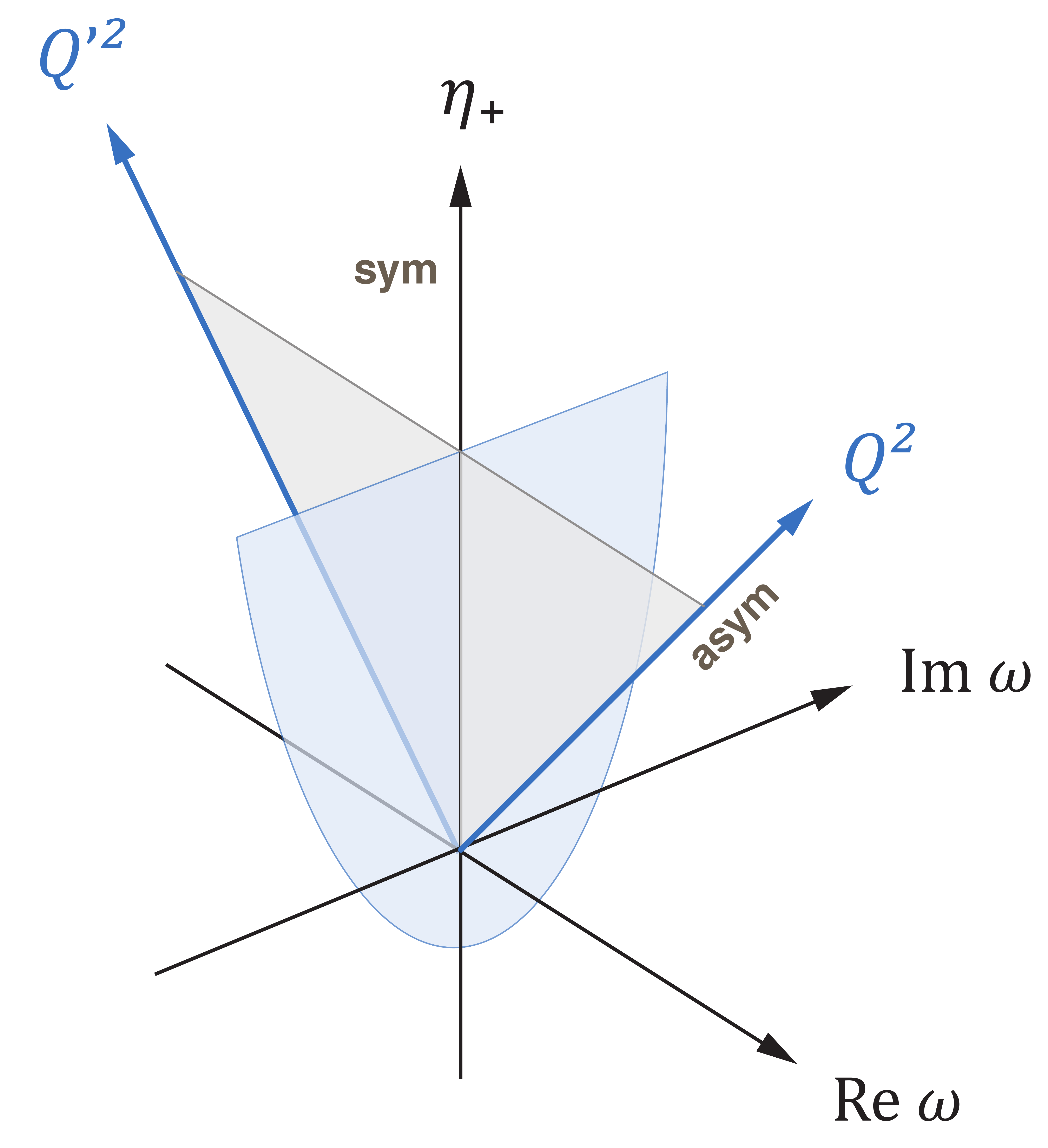}
\caption{Relevant kinematic domain of the transition form factor in the $\pi^0\to e^+ e^-$ decay. The parabola starting at $\eta_+ = -m_\pi^2/4$ is the region that is sampled in the integral.}\label{fig:phasespace-7}
\end{center}
\end{figure}

For arbitrary $t$ the amplitude $\mathcal{A}(t)$ can be defined from the matrix element for the $\pi^0\to e^+ e^-$ decay:
\begin{align}
&  \int \!\! \frac{d^4\Sigma}{(2\pi)^4}\, \Lambda(p_+)\,\gamma^\mu\,S(p+\Sigma)\,\gamma^\nu\,\Lambda(p_-)\,\frac{\Lambda^{\mu\nu}(Q,Q')}{Q^2 \,{Q'}^2} \\
& \quad = \frac{\mathcal{A}(t)}{(4\pi)^2}\,\frac{2im\,\alpha_\text{em}}{\pi f_\pi}\, \Lambda(p_+)\,\gamma_5\,\Lambda(p_-)\,,
\end{align}
where $\Lambda^{\mu\nu}(Q,Q')$ is the $\pi^0\to\gamma\gamma$ transition current from Eq.~\eqref{pigg-current} and $\Lambda(p_\pm)=\frac{1}{2}\,(\mathds{1}+\slashed{p}_\pm/(im))$ is
the positive-energy projector of the lepton.
The kinematics are as discussed in Sec.~\ref{sec:kinematics}; in particular, the averaged photon momentum $\Sigma$ becomes the loop momentum
and therefore the variables in Eq.~\eqref{alt-kinematics} take the values $\sigma>0$ and $Z\in[-1,1]$.
As a consequence, the photon virtualities $Q^2$ and ${Q'}^2$ are tested at complex values close to the symmetric limit as shown in Fig.~\ref{fig:phasespace-7}.
In principle the integral depends on the pion momentum $\Delta$ and the average lepton momentum $p$, but since
the electron and the positron are on-shell with momenta $p_\pm^2=(p\pm \Delta/2)^2=-m^2$ only $t$ remains as an independent variable.

Taking traces yields the following expression for $\mathcal{A}(t)$:
\begin{align}\label{A(t)-1}
             \mathcal{A}(t) &= \frac{1}{2\pi^2 t}\int \! d^4\Sigma\,\frac{(\Sigma\cdot\Delta)^2-\Sigma^2 \Delta^2}{(p+\Sigma)^2+m^2}\,\frac{F(Q^2,{Q'}^2)}{Q^2 \,{Q'}^2}\,.
\end{align}
This integral has poles in the integration domain (which we discuss in more detail in Sec.~\ref{sec:A(t)-direct}) and thus cannot be naively integrated except for the unphysical point $t=\Delta^2/4=0$. A standard way to circumvent the problem uses dispersive methods (see e.g. ~\cite{BERGSTROM1983117,Donoghue:1996kw} ). In that case the imaginary part of the amplitude along its cut at $t<0$ is given by
\begin{align}\label{eq.rare:imag}
& \text{Im} \  \mathcal{A}^{\text{LO}} (t)  = \frac{\pi\,\ln  \gamma(t)}{2 \beta(t)}\, F(0,0)\,,
\end{align}
with $\gamma(t) = (1-\beta(t))/(1+\beta(t))$, which follows from cutting the two photon lines. The imaginary part gives the well-known unitary bound for the branching ratio through the inequality
$\vert \mathcal{A}(t_0) \vert ^2 \geq \vert \text{Im} \mathcal{A}(t_0) \vert^2$:
\begin{align}
R  \geq  \left( \frac{m \alpha_\text{em} }{m_\pi}\right)^2 \frac{\ln^2 \gamma(t_0)}{2\beta(t_0)}    = 4.75 \times 10^{-8}. \nonumber
\end{align}
Using a once-subtracted dispersion relation one then obtains the real part of the amplitude via
\begin{align}\label{eq.rare:re}
     \text{Re} \ \mathcal{A}(t) = \mathcal{A}(0) + \frac{\ln^2 \gamma(t) + \tfrac{1}{3}  \pi^2 + 4\,\text{Li}_2(-\gamma(t))}{4\beta(t)}\,,
\end{align}
where $\text{Li}_2(z)$ is the dilogarithm or Spence function.
In particular, this implies $\text{Re} \ \mathcal{A}(t_0) = \mathcal{A}(0) +31.92(2)$ so that the only unknown left is the constant $\mathcal{A}(0)$.

In fact, $t=0$ is the only point where Eq.~\eqref{A(t)-1} can be integrated directly to yield
\begin{align}
\mathcal{A}(0) = \frac{4}{3}  \int\limits_0^\infty dx \left[ (x-2)\sqrt{1+\frac{1}{x}} - x + \frac{3}{2} \right] F(Q^2,Q^2)\,,
\end{align}
where we temporarily abbreviated $x=Q^2/(4m^2)$. A similar formula can be derived using a Mellin-Barnes representation~\cite{Ghaderi668048,Dorokhov:2008cd,Dorokhov:2009xs},
\begin{align}\label{eq.rare:A0_1}
\mathcal{A}(0) \approx  -\frac{5}{4} + \frac{3}{2} \int\limits_0^{\infty} dx \ln (4x)\, \frac{d}{d x} F(Q^2,Q^2),
\end{align}
which is however only valid to leading order in an expansion in the electron mass. At the point $t=0$
the transition form factor in both cases is evaluated in the symmetric limit of equal photon momenta,
and due to $Q^2 = 4m^2 x$ it is mainly probed at very low $Q^2$ of the order of the electron mass.
Implementing our result for $F(Q^2,{Q'}^2)$, we extract the same value $\mathcal{A}(0)= -21.85(2)$ from both formulas above,
where the error comes from varying the $\xi$ parameter in Eq.~\eqref{fit}.
With Eqs.~(\ref{eq.rare:imag}--\ref{eq.rare:re}) one then arrives at the on-shell value $\mathcal{A}(t_0) = 10.07(4) - 17.45(1)\,i$,
which  corresponds to a branching ratio of
\begin{align}\label{rare-decay-br-1}
B(\pi\rightarrow e^+ e^- )=6.21(3)\times 10^{-8}\,.
\end{align}

\begin{table}[t]
\begin{center}
\begin{tabular}{cl}
\toprule
Collaboration & B($\pi^0\rightarrow e^+ e^-)\, [ 10^{-8}]$ \\\midrule \hline
Experiment \cite{Abouzaid:2006kk,Vasko:2011pi,Husek:2014tna}	&\hspace{1cm} $6.87(36)$\\
Dorokhov et al. \cite{Dorokhov:2007bd,Dorokhov:2009jd} 			&\hspace{1cm} $6.23(9)$\\
Husek et al. \cite{Husek:2015wta}(THS) & \hspace{1cm} $ 6.14(8)$ \\
Masjuan et al. \cite{Masjuan:2015lca} 							&\hspace{1cm} $6.23(5)$\\
Our result (DR)    											    &\hspace{1cm} $6.21(3)$\\
Our result (direct) 											&\hspace{1cm} $6.22(3)$\\
\bottomrule
\end{tabular}
\caption{Our result for the rare decay, obtained either with the dispersion relation (DR) or directly from the contour deformation,
         compared to other theoretical calculations and experiment (after removing the final state radiative corrections). \label{tab:rare}}
\end{center}
\end{table}
Our result is compared to other approaches in Table~\ref{tab:rare}. Whereas our calculation represents
a top-down approach using a well-tested model for the underlying quark-gluon interaction,
Refs.~\cite{Dorokhov:2007bd,Dorokhov:2009jd} use a phenomenological parametrization of the
transition form factor that is adapted to reproduce experimental data from CLEO
together with additional high-energy QCD constraints. A generalization of LMD+V is the two-hadron saturation model (THS) of Ref.~\cite{Husek:2015wta}. The more recent Ref.~\cite{Masjuan:2015lca}
employs a data-driven approach via Pad\'e Theory and Canterbury approximants. All four theoretical results
are in agreement with each other, thus showing consistency between different approaches.
Again, it appears that the decay rate is not overly sensitive to different representations
of the form factor as long as the QCD constraints are satisfied (as guaranteed in all three approaches).
However, we would like to point out that all three calculations rely on dispersion relations and
the Mellin-Barnes representation. Thus the only number that influences the final result is
the constant $\mathcal{A}(0)$. Although \emph{a priori} one would deem the dispersive approach
reliable for this process, it still remains to be checked via a direct calculation.

\subsection{Direct calculation of \texorpdfstring{$\mathcal{A}(t)$}{A(t)}} \label{sec:A(t)-direct}

The integrand in Eq.~\eqref{A(t)-1} has poles for vanishing denominators, i.e., if either of the photons or the intermediate lepton go on-shell. Depending on the value of $t$, this may prohibit a straightforward Euclidean integration. Specifically, for $t \in \mathds{C}$ one can draw a kinematically safe region in the complex $t$ plane where such an integration is possible, and a forbidden region where the poles enter in the integration domain and the integration would produce a wrong result. The latter case would usually be interpreted as a failure of the Wick rotation; however, as we demonstrate below, the Euclidean expression Eq.~\eqref{A(t)-1} is still valid if the poles are treated correctly. Problems of this kind are frequent in Euclidean bound-state calculations and pose limitations, e.g., in computing excited hadrons or form factors for time-like or large space-like arguments~\cite{Eichmann:2016yit} and thus it is desirable to find a general method to deal with them.

In the case of Eq.~\eqref{A(t)-1} it is the unfortunate combination of all three
external momenta being on-shell that complicates the situation.
The analysis in Appendix~\ref{app:sing} shows that the lepton poles lead to a narrow parabola
\begin{align}\label{rare-decay-domain}
(\text{Im}\,t)^2 < 4m^2 \text{Re}\,(-t) \, ,
\end{align}
around the negative (time-like) $t$ axis which is kinematically safe, whereas the photon poles admit a straightforward integration only for real and positive $t$. Taken in combination, the integration is only possible for $t=0$, which leads to the result for $\mathcal{A}(0)$ quoted above.

To analyze the situation for general $t$, we write the integral in hyperspherical variables defined by
\begin{align*}\label{li-2a}
\sigma = \Sigma^2\,, \quad
Z = \frac{\Sigma\cdot\Delta}{2\sqrt{\sigma t}}\,, \quad
Y = \frac{p\cdot \Sigma}{i\sqrt{\sigma}\sqrt{t+m^2}\sqrt{1-Z^2}}\,,
\end{align*}
cf.~Eq.~\eqref{alt-kinematics},
where the process becomes particularly simple in the frame
$\Delta=2\sqrt{t}\,[0,0,0,1]$, $p=i\sqrt{t+m^2}\,[0,0,1,0]$ and
\begin{align}
   \Sigma = \sqrt{\sigma}\left[\begin{array}{l} \sqrt{1-Z^2}\,\sqrt{1-Y^2}\,\sin\psi\\ \sqrt{1-Z^2}\,\sqrt{1-Y^2} \,\cos\psi \\ \sqrt{1-Z^2} \,Y \\ Z \end{array}\right].
\end{align}
The innermost $\psi$ integration is trivial and thus the integral $\mathcal{A}(t)$ takes the form
\begin{align}\label{A(t)-2}
&\mathcal{A}(t) = -\frac{2}{\pi}\int_0^\infty d\sigma\,\sigma^2\int_{-1}^1 dZ\,\frac{(1-Z^2)^{3/2}\,F(Q^2,{Q'}^2)}{(\sigma-t)^2+4\sigma t\,(1-Z^2)} \nonumber \\
& \times\int_{-1}^1 dY\,\frac{1}{\sigma-t +2i\sqrt{\sigma}\sqrt{t+m^2}\sqrt{1-Z^2}\,Y}\,.
\end{align}
The denominator under the $dZ$ integral is equal to $Q^2\,{Q'}^2$ whereas the second denominator corresponds to the lepton pole. The integration in $Y$ can be performed analytically:
\begin{align}\label{rd-log}
   \int_{-1}^1 dY\,\frac{1}{a+ Y} = \ln\,\frac{a+1}{a-1} \, ,
\end{align}
for all $a \in \mathds{C}$ except $-1<a<1$, in which case the logarithm develops a branch cut.

\begin{figure}[t]
\begin{center}
\includegraphics[width=1\linewidth]{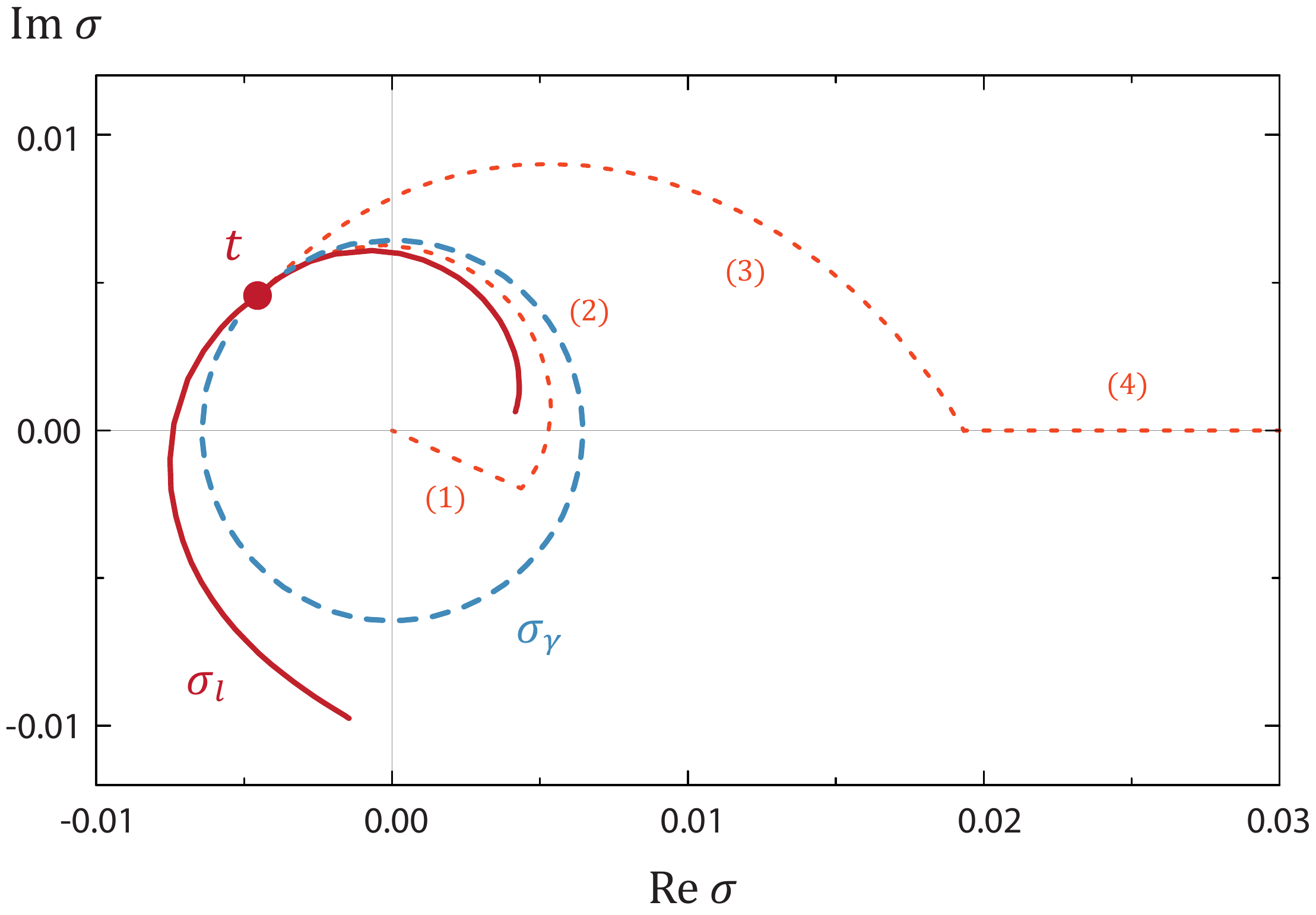}
\caption{Sketch of the overlapping branch cuts in the integrand of $\mathcal{A}(t)$, i.e., the complex $\sigma$ plane, for $t=(-1+i)\,m_\pi^2/4$ and $m=40$ MeV. The cut $\sigma_l$ (solid, red) is generated by the lepton pole and the cut $\sigma_\gamma$ (dashed, blue) by the photon poles; the latter opens at $\sigma=t$ but the former does not. The dotted line shows a possible integration path avoiding all singularities. The units are in GeV$^2$.\label{fig.epemg:rd-cuts}}
\end{center}
\end{figure}

After performing all angular integrations, the conditions $Q^2\,{Q'}^2=0$ and $-1<a<1$ produce poles and cuts
in the complex $\sigma$ plane which are visualized in Fig.~\ref{fig.epemg:rd-cuts}. In principle, the $\sigma$
integration goes from zero to infinity but the `naive' Euclidean' integration path $\sigma \in \mathds{R}_+$
would cross a singularity, hence causing the problems described above. The vanishing denominator for the photons produces a cut along
\begin{align*}
\sigma_\gamma^\pm(Z) = t\,(Z \pm i\sqrt{1-Z^2})^2\,,
\end{align*}
with $-1<Z<1$, which describes a circle with radius $|t|$. The circle opens at $\sigma=t$ since in this case the remaining $(1-Z^2)$ factor in the the denominator of Eq.~\eqref{A(t)-2} cancels with the numerator. On the other hand, the lepton denominator leads to a cut
\begin{align*}
\sigma_l^\pm(Z) = (t+m^2) \left( \sqrt{Z^2-\frac{m^2}{t+m^2}} \pm i \sqrt{1-Z^2}\right)^2,
\end{align*}
which does not open at $\sigma=t$ as this would correspond to $a=0$ in Eq.~\eqref{rd-log}. Instead, it never crosses the arc that passes through $\sigma=-t$ as shown in Fig.~\ref{fig.epemg:rd-cuts}.

We solve the problem by exploiting Cauchy's theorem in finding an integration contour that connects $\sigma=0$ and $\sigma\to\infty$ but never crosses any singularity. Such a possible path is shown in Fig.~\ref{fig.epemg:rd-cuts}: (1) it departs from the origin in the direction opposite to $t$, (2) then navigates between the cuts until it reaches the point $\sigma=t$, (3) returns to the positive real axis at a value $\sigma > |t|$, (4) and from there proceeds to the numerical cutoff at $\sigma\to\infty$. This strategy ensures that for each $\sigma$ along the path the integrand in $Z$ is free of any singularities.

In practice, however, the problem is made worse by the small lepton mass $m \ll m_\pi$: in that case both cuts essentially describe the same circle but open on opposite sides, $\sigma=t$ and $\sigma=-t$, so that the path along segment (2) proceeds along a narrow ridge between the two cuts. As a consequence, the integrand in $Z$ is sharply peaked which requires an enormous number of grid points to obtain stable results.
To this end we have optimized the procedure so that path (2) is always equally spaced between the cuts, thus minimizing the numerical error. In addition, we implement an adaptive integration in $-1 < Z < 1$ that accumulates the grid points according to the nearest singularities in the complex $Z$ plane, which are determined by
solving $\sigma_\gamma^\pm(Z)$ and $\sigma_l^\pm(Z)$ for $Z$. In that way we gain a factor of $\sim 10^3$ in CPU time while maintaining the same numerical accuracy.

\begin{figure}[t]
\begin{center}
\includegraphics[width=0.7\linewidth]{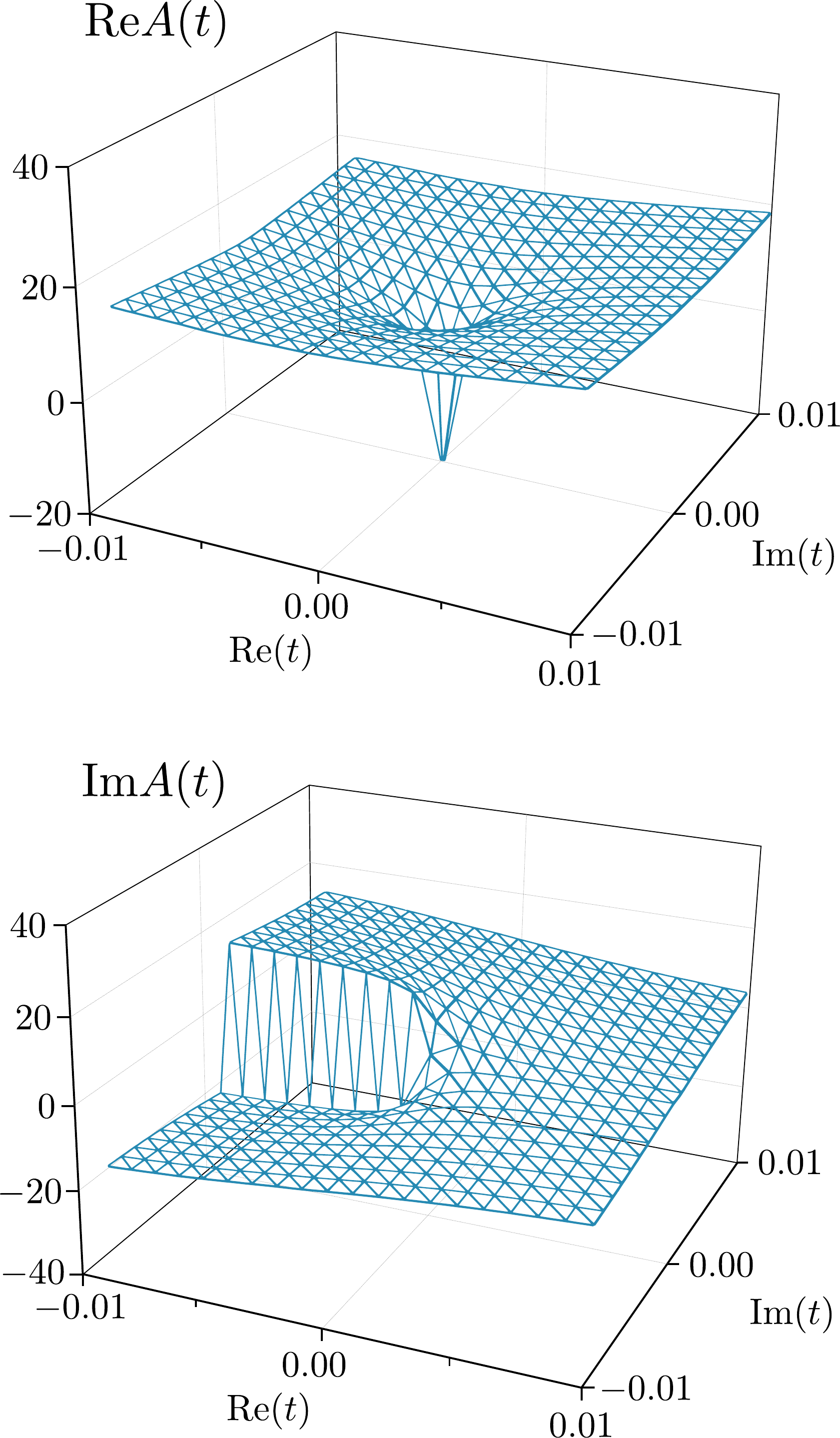}
\caption{Result for $\mathcal{A}(t)$ in the complex $t$ plane. The units for $t$ are GeV$^2$ and $\mathcal{A}(t)$ is dimensionless. The on-shell pion point is $t_0 = -m_\pi^2/4 \approx -0.005$ GeV$^2$.\label{fig.epemg:At}}
\end{center}
\end{figure}

As a result, we are in principle able to determine $\mathcal{A}(t)$ in the whole complex plane, which is
shown in Fig.~\ref{fig.epemg:At} in the momentum region relevant for the $\pi^0\to e^+ e^-$ decay.
The real part is sharply peaked at the origin $t=0$, whereas the imaginary part develops the expected branch cut on the timelike axis.
The resulting on-shell value is $\mathcal{A}(t_0) = 10.10(3)-17.45(1)\,i$, where the error reflects the uncertainty in the TFF discussed below Eq.~\eqref{fit-parameters}.
The corresponding branching ratio
\begin{align}\label{rare-decay-br-2}
B(\pi^0\rightarrow e^+ e^- )=6.22(3)\times 10^{-8} \, ,
\end{align}
nicely agrees with the result in Eq.~\eqref{rare-decay-br-1} obtained via dispersion relations.

We have thus established a fast and efficient numerical method to calculate Euclidean integrals,
which is applicable even in cases where a naive Wick rotation fails.
It can be generalized to arbitrary integrals as long as the singularity structure of the integrand is known,
which is not restricted to real poles but accommodates complex poles or cuts as well.
Perhaps unsurprisingly, this demonstrates
that both Euclidean and Minkowski descriptions are completely equivalent as long as the singularities in the integrand are treated correctly:
the correct Euclidean expression, such as Eq.~\eqref{A(t)-2}, is the one that connects zero with infinity without crossing any singularities.

Exploratory calculations using contour deformations have been performed, e.g., in determining quark, gluon and ghost propagators in the complex plane~\cite{Maris:1995ns,Alkofer:2003jj,Eichmann:2009zx,Strauss:2012dg,Windisch:2012sz}.
In those cases the singularity structure of the integrands is usually less intertwined. In turn, one has to deal with integral equations and thus self-consistent problems where
the singularities that are dynamically generated by the integration enter again into the integrand and must be accounted for as well.
In addition to other methods to calculate propagators in the complex plane~\cite{Fischer:2005en,Fischer:2008sp,Krassnigg:2009gd,Rojas:2014aka}, or a direct inclusion of residues as done for example in Ref.~\cite{Oettel:1999gc},
contour deformation methods may become an attractive tool for overcoming several long-standing obstacles in the Dyson-Schwinger/Bethe-Salpeter approach, e.g.,
the calculation of (highly) excited states, form factors at timelike or large spacelike momentum transfer, or the treatment of genuine resonances.

\section{Conclusions}\label{sec5}

In this work we determined the branching ratios of various leptonic decays of the neutral pion. The common central
element in these calculations is the pion transition form factor. We calculated its momentum dependence in a
Dyson-Schwinger and Bethe-Salpeter framework employing an underlying quark-gluon interaction that has been
successful elsewhere in describing a range of different static and dynamic meson and baryon
properties, see e.g.~\cite{Maris:2003vk,Eichmann:2016yit} for reviews. The resulting form factor dynamically generates vector meson dominance and can be shown to comply with the perturbative scaling behavior at large space-like momenta in the symmetric limit, whereas the coefficient of the scaling limit is modified 
in the asymmetric limit~\cite{Eichmann:2017wil}.

Due to the kinematic constraints imposed by Dalitz decays -- as is well-known --
only a very small range of time-like momenta around the zero-momentum limit controlled by the anomaly is probed.
Consequently, our results for these pion decays very much resemble those of other theoretical
approaches that use, e.g., vector meson dominance models for the form factor supplemented with constraints from the
anomaly. The results are subsequently compatible with experiment.

For the rare decay the form factor is probed in the space-like region and we have been able to confirm theoretical calculations using dispersion relations,
however through a direct calculation using contour deformations. Thus our result for the rare decay $\pi^0 \rightarrow e^+ e^-$ still leaves a 2$\sigma$ discrepancy
between theory and experiment.

\section{Acknowledgements}
We thank R. Alkofer, S. Leupold, A. Szczepaniak and M.~T.~Hansen for enlightening discussions.
This work was supported by the DFG collaborative research centre TR 16, the BMBF grant 05H15RGKBA,
the DFG Project No. FI 970/11-1, the FCT Investigator Grant IF/00898/2015,
the GSI Helmholtzzentrum fuer Schwerionenforschung, and by the Helmholtz International Center for FAIR.

\begin{figure*}[t]
\begin{center}
\includegraphics[width=1\textwidth]{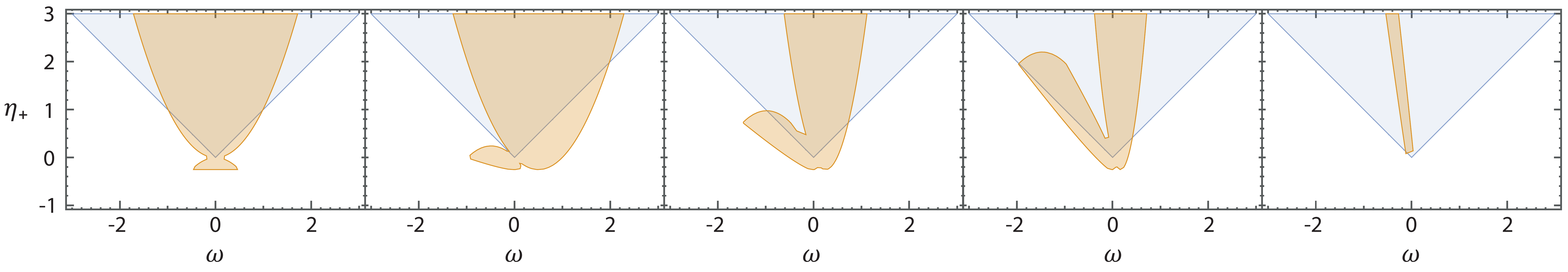}
\caption{Kinematically accessible regions in the $(\omega,\eta_+)$ plane for different frames; the units are in GeV$^2$.
From left to right: $\alpha=0$, $\nicefrac{1}{2}$, $1$, $\nicefrac{3}{2}$, $8$.
The (blue) triangles are merely drawn for guidance and show the spacelike domain ($Q^2>0$, ${Q'}^2>0$).
}\label{fig:sing}
\end{center}
            \end{figure*}

\appendix

\section{Singularity restrictions}\label{app:sing}

Here we return to the problem mentioned in Section~\ref{sec:tff-result}, namely, to find the kinematic regions
where a Euclidean integral with singularities in the integrand can be calculated directly without any contour deformations.
For generality, consider a Lorentz-invariant integral
\begin{align}
\mathcal{I}(p_1, \dots p_n) = \int d^4k_1 \int d^4k_2\, \dots f(q^2) \dots
\end{align}
It depends on a collection of external momenta $p_i^\mu$ and one integrates over the loop momenta $k_j^\mu$. The integrand consists of Lorentz-invariant functions such as $f(q^2)$, where $q^\mu$ is a linear combination of the external and loop momenta. The loop momenta $k_j^\mu$ are always real; however, if an external momentum is time-like ($p_i^2<0$) it will inject imaginary components into $q^\mu$, which is therefore a complex four-vector. Splitting $q^\mu$ into its real and imaginary parts, where $A$, $B\in\mathds{R}$ and $e_\mu$, $e'_\mu$ are Euclidean unit vectors, yields:
\begin{align*}
q_\mu = A\,e_\mu + i B\,e'_\mu \; \Rightarrow \; q^2 = A^2 - B^2 +2i  A B\,(e\cdot e') \,.
\end{align*}
Because $e\cdot e' \in [-1,1]$, the Lorentz invariant $q^2$ is tested inside a parabola $(A \pm iB)^2$ with apex $-B^2$ on the real axis; for $B=0$ it becomes the positive real axis. In other words, if $q^\mu$ has imaginary components then $q^2$ is sampled inside a complex parabola.

Now, if $f(q^2)$ has singularities in the complex plane, the corresponding parabola passing through the first singularity (e.g., a real pole, complex conjugate poles, or the onset of a cut) defines the `contour mass' $q^2=-m_p^2$. The kinematically safe region is then subject to the restriction
\begin{align}
-B^2 > -m_p^2 \quad \Leftrightarrow \quad \left[ \text{Im}\,q \right]^2 = B^2 < m_p^2\,.
\end{align}
Because the imaginary part of $q^\mu$ can only come from the external momenta $p_i^\mu$, this imposes restrictions on the domain of the external Lorentz invariants.

We specifically consider the transition matrix element in Eq.~\eqref{eqn:PseudoScalarFormFactor}, where the quark momenta in the loop are given by $k \pm \Delta/2$ and $k + \Sigma$. The dressing functions of the quark propagator in Eq.~\eqref{quarkprop},
\begin{align*}
\frac{Z_f(p^2)}{p^2  + M^2(p^2)} \quad \text{and} \quad \frac{Z_f(p^2)\,M(p^2)}{p^2  + M^2(p^2)}\,,
\end{align*}
develop a certain singularity structure in the complex plane. In a rainbow-ladder truncation the nearest singularities are complex conjugate poles
with a typical contour mass $m_p \sim 0.5$ GeV for light quarks (see~\cite{Windisch:2016iud} for a detailed investigation). The loop momentum $k$ is always real and thus one arrives at the conditions
\begin{align}
\left( \text{Im}\,\frac{\Delta}{2}\right)^2 < m_p^2  \quad \text{and} \quad
\left( \text{Im}\,\Sigma\right)^2 < m_p^2\,.
\end{align}

In general these restrictions depend on the chosen frame for $\Sigma$ and $\Delta$.
The components of $\Sigma$ and $\Delta$ can be arranged arbitrarily as long as the Lorentz invariants
defined in Eqs.~(\ref{li-2}--\ref{alt-kinematics}),
$\Sigma^2 = \sigma = \eta_+ + m_\pi^2/4$, $\Delta^2 = -m_\pi^2$ and $\Sigma\cdot\Delta = \omega$, remain unchanged.
For any possible choice, however, one can find a linear combination $\Sigma + \alpha\,\Delta$ that  has a four-component only,
with an arbitrary parameter $\alpha \in \mathds{R}$. The general arrangement satisfying these constraints is
\begin{align*}\label{general-frame}
\Delta = \frac{1}{\mathcal{N}}\left[\begin{array}{c} 0 \\ 0 \\ \!-i\sqrt{\sigma m_\pi^2+\omega^2} \\
\omega-\alpha m_\pi^2 \end{array}\right]\!\!, \quad
\Sigma = \frac{1}{\mathcal{N}}\left[\begin{array}{c} 0 \\ 0 \\ i\alpha\sqrt{\sigma m_\pi^2+\omega^2} \\
\sigma + \alpha \omega \end{array}\right]
\end{align*}
with $\mathcal{N}=\sqrt{\sigma+2\alpha \omega - \alpha^2 m_\pi^2}$.

Take for example the pion's rest frame, which corresponds to  $\alpha\to\infty$:
\begin{align*}
\Delta = \left[\begin{array}{c} 0 \\ 0 \\ 0 \\ im_\pi \end{array}\right]\!\!, \quad
\Sigma = \left[\begin{array}{c} 0 \\ 0 \\ \!\sqrt{\frac{\omega^2}{m_\pi^2} + \sigma} \\
\frac{\omega}{im_\pi} \end{array}\right]
= \sqrt{\sigma} \left[\begin{array}{c} 0 \\ 0 \\ \!\sqrt{1-Z^2} \\  Z \end{array}\right].
\end{align*}
If $\sigma>0$ and $Z \in [-1,1]$, $\Sigma$ is real and the imaginary part only comes from $\Delta$, so the resulting condition $m_\pi < 2m_p$ is always satisfied. However, the situation is different in the quadrant $\eta_+>0$, $|\omega| < \eta_+$ shown in Fig.~\ref{fig:phasespace-1}, because in this case $Z$ is imaginary. Here both $\Delta$ and $\Sigma$ have imaginary four-components and the resulting condition becomes
$\left|\omega\right| < m_\pi m_p$, which defines a narrow strip around the symmetric limit $\omega=0$.

The arbitrariness of $\alpha$ can be exploited to optimize the frame, i.e. to reach kinematic regions for the form factor that are not accessible in the pion rest frame. The resulting domains are plotted in Fig.~\ref{fig:sing} for different values of $\alpha$. The leftmost plot shows $\alpha=0$ and the rightmost plot $\alpha=8$; for $\alpha\to\infty$ one recovers the pion rest frame. For example, with $\alpha=0$ the momentum $\Sigma$ is the one with a four-component only, whereas $\Delta$ has an imaginary three-component and, for $\sigma>0$, leads to the condition
\begin{align}
\omega^2 < (4m_p^2-m_\pi^2)\left( \eta_+ + \frac{m_\pi^2}{4}\right).
\end{align}

For the singly-virtual form factor $F(Q^2,0)$ the optimal choice is $\alpha=1/2$ (second plot in Fig.~\ref{fig:sing}). In that case it is the photon momentum $Q = \Sigma + \Delta/2$ that has a four-component (resembling the Breit frame in elastic form factor calculations). The denominator $\mathcal{N} = \sqrt{\eta_++\omega}$ is always real if $\eta_+>0$ and $|\omega| < \eta_+$, and therefore the three-components of $\Sigma$ and $\Delta$ are imaginary. The resulting condition is
\begin{align}
m_\pi^2\, \eta_+ + \frac{m_\pi^4}{4} + \omega^2 < 4m_p^2 \,(\eta_++\omega)\,.
\end{align}
This region crosses the line $\eta_+=\omega$ at
\begin{align}
\eta_+=\omega=\frac{Q^2}{2} = 4m_p^2 \left( 1 - \frac{\varepsilon}{2} + \sqrt{1-\varepsilon}\right) \approx 8m_p^2\,,
\end{align}
with $\varepsilon=m_\pi^2/(4m_p^2)$. Hence, in the asymmetric limit the form factor can be calculated up to $Q^2_\text{max} \approx 16\,m_p^2 \approx 4$~GeV$^2$ without crossing any quark singularities.

As a second example we consider the integral $\mathcal{A}(t)$ in Eq.~\eqref{A(t)-1} which describes the $\pi^0\to e^+ e^-$ decay. In that case the external momenta are $\Delta$ and $p$, with Lorentz invariants $\Delta^2 = 4t$, $p^2=-(t+m^2)$ and $p\cdot\Delta=0$, whereas $\Sigma$ is the real loop momentum. The internal photon momenta are $\Sigma \pm \Delta/2$ and the lepton momentum is $\Sigma+p$, so the singularity conditions become
\begin{align}
\left( \text{Im}\,\frac{\Delta}{2}\right)^2 < 0  \quad \text{and} \quad
\left( \text{Im}\,p\right)^2 < m^2\,,
\end{align}
where $m$ is the lepton mass. The analogous arrangement in the general frame is
\begin{align*}\label{general-frame-2}
\Delta = \frac{1}{\mathcal{N}}\left[\begin{array}{c} 0 \\ 0 \\ \!-2i\sqrt{t}\sqrt{t+m^2} \\ 4\alpha t \end{array}\right]\!\!, \,\,
p = \frac{1}{\mathcal{N}}\left[\begin{array}{c} 0 \\ 0 \\ 2i\alpha\sqrt{t}\sqrt{t+m^2} \\ -(t+m^2) \end{array}\right],
\end{align*}
with $\mathcal{N}=\sqrt{-(t+m^2)+4\alpha^2 t}$. In this case the maximal domains correspond to the limits
$\alpha=0$ or $\alpha\to\infty$ and one arrives at the condition $(\text{Im}\,t)^2 < 4m^2\, \text{Re}\,(-t)$
from the lepton pole and $t>0$ from the photon poles, which are quoted in Eq.~\eqref{rare-decay-domain}. Although in principle these regions are disjoint,
the analysis of Eq.~\eqref{A(t)-2} shows that the point $t=0$ contains an integrable singularity and thus $\mathcal{A}(0)$ is well-defined if $m>0$.

\section{Four-body phase space}\label{app:4bphase}
In this appendix we work out the four-body phase-space integral $d\Phi_4$ that enters in the $\pi^0\to e^+ e^- e^+ e^-$ decay of Eq.~\eqref{four-body-decay-rate}.
The decay width of a particle with momentum $P$ and mass $M$ decaying into $n$ daughter particles with momenta $p_i$ and masses $m_i$ is given by
\begin{align}
\Gamma (P\rightarrow  p_i) = \frac{1}{S}\,\frac{1}{2 M}\int d \Phi_n \, | \mathcal{M}|^2   \, ,
\end{align}
with $S$ the symmetry factor, $| \mathcal{M}|^2$ the spin summed and squared matrix element, and $d \Phi_n$ the phase-space integral for an $n$-particle final state given by
\begin{align}
d \Phi_n =   (2\pi)^4 \delta^4\left(P- \sum_{i=1}^n p_i \right) \prod_{i=1}^{n} \frac{d^3 \textbf{p}_i}{(2\pi)^3 \,2 E_i} \, .
\end{align}

Following Ref.~\cite{Byckling:1971vca},  we rewrite the integration in terms of invariant mass variables.
For $n=4$ one obtains
\begin{equation}\label{app.eq:phasespace}
d \Phi_4 = \frac{1}{(2\pi)^8} \frac{\pi^2}{32 M^2} \, \frac{ d s_{12}\, d s_{34} \,d s_{124} \,d s _{134}  \,d s_{14} }{\sqrt{-\Delta_{(4)}}}
\end{equation}
where, for degenerate decay products with $m_i=m$, the two- and three-particle Mandelstam variables read
\begin{equation}\label{app.eq:sij} 
\begin{split}
   s_{ij} &= -(p_i+p_j)^2 = 2m^2 - 2p_i\cdot p_j\,, \\
   s_{ijk} &= -(p_i+p_j+p_k)^2 = s_{ij} + s_{ik} + s_{jk} - 3m^2
\end{split}
\end{equation}
and the four-dimensional Gram determinant $\Delta_{(4)}$ contains all possible dot products of four-vectors:
\begin{equation} \label{appen:eq.delta4} \renewcommand{\arraystretch}{1.2}
    \Delta_{(4)} = \det\left[ \begin{array}{cccc}
        -m^2 & p_1\cdot p_2 & p_1\cdot p_3 & p_1\cdot p_4 \\
p_1\cdot p_2 &         -m^2 & p_2\cdot p_3 & p_2\cdot p_4 \\
p_1\cdot p_3 & p_2\cdot p_3 &         -m^2 & p_3\cdot p_4 \\
p_1\cdot p_4 & p_2\cdot p_4 & p_3\cdot p_4 &  -m^2
                        \end{array}\right].
\end{equation}
In contrast to \cite{Escribano:2015vjz,Byckling:1971vca,Nyborg:1965zz} we employ a Euclidean signature, however
with $s_{ij}$ and $s_{ijk}$ defined such that they have the same meaning in Minkowski and Euclidean conventions.
To work with the invariant mass variables of Eq.~\eqref{app.eq:phasespace}, one replaces the $p_i\cdot p_j$ in the Gram determinant according to Eq.~\eqref{app.eq:sij} together with
\begin{align}
\sum_{i<j} s_{ij} = -(M^2 + 8m^2) \, .
\end{align}

The physical region of integration is bounded by the surface $\Delta_{(4)}=0$.
Following the derivation of Refs.~\cite{Escribano:2015vjz,byers_physical_1964,Nyborg:1965zz},
we impose this relation on the invariant mass variables
and begin by solving $\Delta_{(4)}=0$ for $s_{14}$. It yields
\begin{equation}\label{ap.eq.s14b}
 s_{14}^{\pm} = \frac{b \pm 2 \sqrt{G(s_{124},s_{34},s_{12})\,G(s_{134},s_{12},s_{34})}}{\lambda(s_{12},s_{34},M^2)}\, ,
\end{equation}
where $\lambda(u,v,w) =  u^2+v^2+w^2-2uv-2uw-2vw$ is the K\"allen function, the $G$ functions are given by
\begin{equation*}
\begin{split}
  &G(u,v,w) = m^2(w-M^2)^2 \\
  & +v\left[(u-m^2)^2-(u+m^2)(w+M^2)+w M^2+uv\right],
\end{split}
\end{equation*}
and $b$ is the coefficient of $-8\Delta_{(4)}$ linear in $s_{14}$:
\begin{equation*}
\begin{split}
  b &= G(s_{124},s_{34},s_{12}) + G(s_{134},s_{12},s_{34}) \\
    &+ M^2 c d - (c+d)(s_{12}\, d + s_{34}\, c)
\end{split}
\end{equation*}
with $c=s_{124}-m^2$ and $d=s_{134}-m^2$.

The regions of the $s_{124}$ and $s_{134}$ integrations are bounded by the surfaces satisfying $s_{14}^+= s_{14}^-$, which
is fulfilled when either of the $G$ functions in Eq.~\eqref{ap.eq.s14b} vanishes $G(s_{124},s_{34},s_{12})  = 0$ or $G(s_{134},s_{12},s_{34}) =0$.
Solving $G(u,v,w)=0$ for $u$ yields
\begin{equation*}
  s_{14}^\pm = \frac{w-v+M^2+2m^2}{2} \pm \frac{\sqrt{v-4m^2}\sqrt{\lambda(v,w,M^2)}}{2\sqrt{v}}
\end{equation*}
and thus determines the integration boundaries $s_{124}^\pm$ and $s_{134}^\pm$ as functions of the two dilepton invariant masses $s_{12}$ and $s_{34}$.

Finally, $s_{34}$ and $s_{12}$ range from the threshold at $4m^2$ to $(M-\sqrt{s_{12}})^2$ and
$(M-2m)^2$, respectively; here the ordering is arbitrary and could also be exchanged.

A valuable check when rewriting the phase space integral in different coordinates is the massless limit. For massless daughter particles the $n$-body phase space is given by
\begin{align}
\Phi_n =\frac{1}{2\,(4\pi)^{2n-3}} \frac{M^{2n-4}}{\Gamma(n)\,\Gamma(n-1)} \, .
\end{align}
Integrating over the phase space volume only, as given in  Eq.~\eqref{app.eq:phasespace} with the borders as suggested for $m=0$, reproduces the limit exactly as it should.

The final decay rate for the decay of the neutral pion ($M=m_\pi$) into two dileptons ($m=m_e$) is then given by
\begin{widetext}
\begin{align}\label{finalphasespace}
 \Gamma_{\pi^0 \rightarrow 2e^+ 2e^-} = \frac{1}{2^{16}\,\pi^6\, m_\pi^3} \!\!\!\!
 \int\limits_{4m_e^2}^{(m_\pi-2m_e)^2} \!\!\!\!\!\!\!\! d s_{12}
 \int\limits_{4m_e^2}^{(m_\pi-\sqrt{s_{12}})^2}  \!\!\!\!\!\!\!\! d s_{34} \;\;
 \int\limits_{s_{124}^-}^{s_{124}^+} d s_{124}
 \int\limits_{s_{134}^-}^{s_{134}^+} d s_{134}
 \int\limits_{s_{14}^-}^{s_{14}^+} d s_{14}
 \,\frac{| \mathcal{M}|^2}{\sqrt{-\Delta_{(4)}}} \,.
\end{align}
\end{widetext}
\section*{References}
\bibliography{baryonspionff}

\end{document}